# The polymorphous nature of cubic halide perovskites

Xin-Gang Zhao[1], Gustavo M. Dalpian[1,2], Zhi Wang[1] and Alex Zunger[1]

[1]*Energy Institute, University of Colorado, Boulder, Colorado 80309*
[2]*Centro de Ciências Naturais e Humanas, Universidade Federal do ABC, 09210-580, Santo André, SP, Brazil*

**ABSTRACT:** Many common crystal structures can be described by a single (or very few) repeated structural motifs ("monomorphous structures") such as octahedron in cubic halide perovskites. Interestingly, recent accumulated evidence suggests that electronic structure calculations based on such macroscopically averaged monomorphous cubic (Pm-3m) halide perovskites obtained from X-ray diffraction, show intriguing deviations from experiment. These include systematically too small band-gaps, dielectric constants dominated by the electronic, negative mixing enthalpy of alloys, and significant deviations from the measured pair distribution function. We show here that a minimization of the systems T=0 internal energy via density functional theory reveals a *distribution* of different low-symmetry local motifs, including tilting, rotations and B-atom displacements ("polymorphous networks"). This is found only if one allows for larger-than-minimal cell size that does not geometrically exclude low symmetry motifs. As the (super) cell size increases, the energy is lowered relative to the monomorphous cell, and stabilizes after ~32 formula units ($\geq$160 atoms) are included. Being a result of non-thermal energy minimization in the internal energy without entropy, this correlated set of displacements must represent the intrinsic geometry preferred by the underlying chemical bonding (lone pair bonding), and as such has a different origin than the normal, dynamic thermal disorder modeled by molecular dynamics. Indeed, the polymorphous network, not the monomorphous ansatz, is the kernel structure from which high temperature thermal agitation develops. The emerging physical picture is that the polymorphous network has an average structure with high-symmetry, yet the local structural motifs have low symmetries. We find that, compared with monomorphous counterparts, the polymorphous networks have significantly lower predicted total energies, larger band gaps and ionic dominated dielectric constants, and agrees much more closely with the observed pair distribution functions. An analogous polymorphous situation is found in the paraelectric phase of a few cubic oxide perovskites where local polarization takes the role of local displacements in halide perovskites, and in the paramagnetic phases of a few 3d oxides where the local spin configuration takes that role.

## I. Introduction

ABX$_3$ compounds often appear (when X=Oxygen) as ferroelectrics, Mott insulators and transparent conductors, and (when X=halogen) as solar photovoltaic absorbers [1,2]. They generally have at low temperatures the low-symmetry monoclinic, orthorhombic or tetragonal ground state structures, and, at higher temperatures the high-symmetry cubic (Pm-3m) structure. The high temperature cubic phase in halide perovskites ABX$_3$ (A=Cs, MA, FA; B=Sn, Pb; X=Cl, Br, I) and in the oxide perovskites SrTiO$_3$, SrVO$_3$ or BaZrO$_3$, are described by X-Ray diffraction structure determination as monomorphous structures, *i.e.*, they have a single octahedral structural motif represented crystallographically by a small, repeated unit cell, as illustrated in Fig. 1a.

We find that removing the standard restriction to such a minimal unit cell size in structural optimization of the internal energy part $H_{cubic}$ of the free energy $H_{cubic} - TS$ leads in many cubic perovskites to the formation of a '*polymorphous network*' [3], manifesting a distribution of different tilt angles and different B-atom displacement in different octahedra, illustrated in Fig. 1(b, c). This distribution emerges already from the (density functional) minimization of the static, $T=0$ internal energy $H_{cubic}$ of a large supercell (see $T=0$ region on the left-hand side of Fig 1d), constrained to have the global cubic lattice vectors. This a-thermal distribution of correlated set of displacement represents the preference of the low temperature chemical bonding (such as lone pair orbitals of Sn$^{2+}$ and Pb$^{2+}$ encouraging stereochemically off-center motions [4]), before thermal agitation sets in. Such static displacements are different than the entropic thermal driving force (illustrated by the finite $T$ range in the right-hand side of Fig 1d), often described by molecular dynamics [5–13]. When the dynamic thermal displacements are averaged over snapshots, they yield the ideal, undisplaced structure, whereas the static displacements obtained by minimizing $H_{cubic}$ (T=0) are inherently non-thermal. Indeed, the polymorphous network (rather than the monomorphous assumption) constitutes the kernel structure from which thermal disorder emerges at elevated temperatures. The polymorphous displacements are different from the single sharp monomorphous values of these deformation parameters (shown as vertical blue lines in Fig. 1c), or from the periodically repeated ordered double-potential well models that address anharmonic polar fluctuations [14–16]. The existence of such a polymorphous distribution is easy to miss using standard energy minimization protocols (such as those based on following gradients to the nearest local minimum) but is revealed once one initially applies a random atomic displacement ("nudge") off the cubic sites and explores lower symmetries in the minimization process.

The emerging physical picture is that the phase seen by XRD as cubic has, in fact local structural motifs with low symmetries which emerge from the intrinsic low *T* chemical bonding pattern. This is different from ordinary inorganic compounds where reduced local symmetries emerge just from thermal disorder at elevated temperatures. The extensively discussed, single formula unit, cubic (Pm-3m) structure of halide perovskites [4,5,15,17–36] does not really exist, except as a macroscopically averaged fictitious structural model. Because X-ray diffraction has a rather long coherence length, such polymorphous systems were often fit in structure refinement models [16,37] by a macroscopically averaged (fictitious monomorphous) cubic (Pm-3m) unit cells. Standard electronic structure calculations [4,15,17–28] that use input structures directly from crystal databases have often modeled the properties of the system (band gaps, absorption spectra, thermodynamic stability, alloy mixing enthalpies) as the property $<P>=P(S_0)$ of the reported macroscopically averaged monomorphous structure [38,39] $S_0$ rather than the average $P_{obs}=\Sigma P(S_i)$ of the properties $\{P(S_i)\}$ of the individual, low symmetry microscopic configurations $\{S_i ; i=1, N\}$. One expects, as illustrated in this paper for a range of properties, that the properties $P(S_0)$ of the unphysical high symmetry cubic structure $S_0$ would differ significantly from the properties $\{P(S_i)\}$ of broken symmetry cubic structures or their average $P_{obs}$.

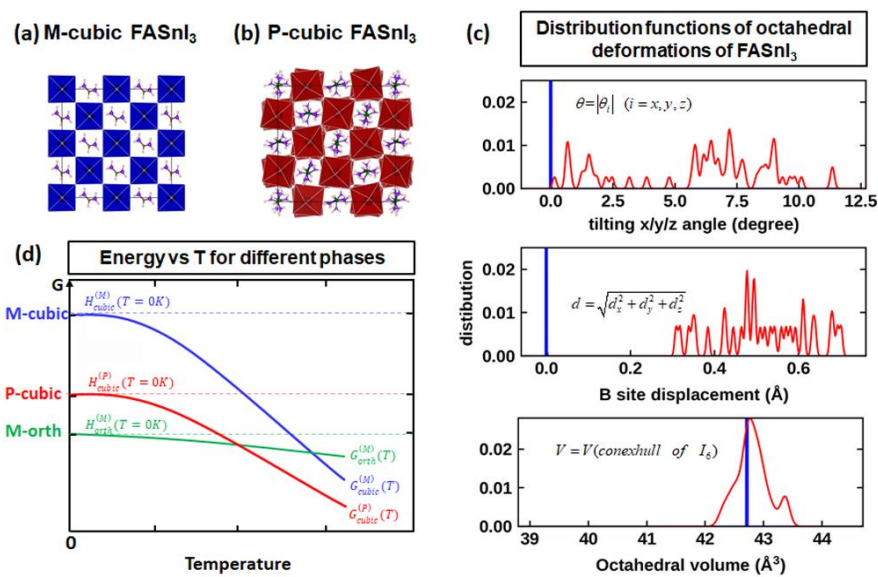

**Figure 1.** Structures of monomorphous cubic (M-cubic) (a) and polymorphous cubic (P-cubic) (b) halide perovskite FASnI$_3$, and the distribution functions (c) of various octahedral deformations obtained by minimization of the internal energy in Density Functional supercell calculations. The panels in (c) show the distribution functions of octahedral titling along x/y/z (top panel), B site off-center displacement (middle panel), and the volumes of individual octahedra (bottom panel). Red and blue solid lines refer to Polymorphous and monomorphous networks, respectively. The supercell structures were optimized by keeping fixed the cubic cell shape. The panel (d) is the schematics of the energies (enthalpy *H*, Gibbs free energy *G*) as function of temperature for cubic monomorphous phase (blue lines), cubic polymorphous (red lines) phase, and ground state orthorhombic (green lines) phase. The total energy lowering $\delta E^{(P-M)}$ for the polymorphous cubic phase with respect to the cubic monomorphous phase was depicted with double arrow.

We find here that the use of the cubic polymorphous network in electronic structure calculations yields significantly improved results relative to the monomorphous assumption, solving many of the outstanding inconsistencies noted previously. These include (i) The polymorphous network has a significant lowering of calculated total energies, (by ~70-150 meV/f.u) (ii) The DFT-calculated Pair Distribution Function (PDF), which probes the local environment, agrees much better with the measurements. The average structure differs from the actual low-symmetry local structure revealed by neutron diffraction studies [8], by local probes such as PDF [40,41]. (iii) Up to 300% larger calculated band gaps. This intrinsic increase is much larger than the additional temperature-induced increase obtained in literature calculated molecular dynamics [6,42]. Thus, (iv) the band gap renormalization energy (~200 meV) is now closer to experiment relative to the values suggested in Molecular Dynamics (MD) calculations (390-640 meV), taken with respect to the band gap of the monomorphous model (v) Use of the polymorphous structure leads to the reversal of the predicted sign of the mixing enthalpies of the solid solutions from negative (ordering-like; not seen experimentally) to positive (experimentally observed phase-separating), removing the previously existing qualitative conflict with experiment. (vi) Remarkably, despite the existence of a distribution of motifs, the calculated band structure (unfolded to the primitive Brillion zone from the supercell) shows sharp band edge states and a correspondingly fast rise of the absorption spectrum, in agreement with experiment and consistent with the fact that the calculated displacements are correlated. This is different than the naïve expectation based on ordinary disorder models where disorder is expected to fill–in the band gap region by localized states, and lead to a broad and slow rising absorption tail. Finally, relative to the monomorphous case, (vii) polymorphous networks have a much larger (by ~50%) calculated dielectric constant, where the ionic contribution now dominates the electronic contribution as expected from near ferroelectrics.

## II. Ambiguities and contradictions regarding the assumed monomorphous cubic structure

The nominally cubic phase of halide perovskites is the leading candidate for high efficiency solar cells, (reaching 24% as tandem [43]), enabled by band gaps in the solar range of 1-2 eV, sharp absorption edges, and long carrier lifetimes. The question addressed here is what is the actual atomic arrangement of this all-important cubic phase [40,44–46]. A number of observations cast doubt on its traditional description as a *monomorphous cubic perovskite*.

***Structure refinement based on periodic monomorphous cells does not lead to a satisfactory fit to the data***: X-ray structural studies [46] have fitted intensities to a single formula unit Pm-3m cubic structure. Local probes such as Pair Distribution Function (PDF) provide a better rendering of the structure. The PDF of the monomorphous model calculated with DFT (method details are described in Appendix-A, S-1) is shown in Fig. 2b for MAPbI$_3$ and gives qualitative disagreements, especially for the peaks P3a (describing Pb-Pb and Pb-I distances) and P3b (describing Pb-I). A good PDF fit to experiment within a monomorphous ansatz was possible in the literature only when assuming model parameters (such as anisotropic, non-spherical atoms and displaced X sites [15,44,46]) that in the single formula unit model cannot possibly describe a periodic crystalline network. Consequently, such literature fitted non-periodic model with effective parameters [40] cannot be used in a Schrödinger equation to predict electronic structure and optical properties. Fig. S-2 in Appendix B compares the PDF of double well model (without fitting parameters, unlike Ref [40]) with the present results.

*Gap anomaly*: Whereas the measured gaps of cubic FASnI$_3$ is ~100 meV *larger* than that of orthorhombic CsSnI$_3$, calculations using the monomorphous cubic (M-cubic) structure (Table I and Fig. S-3 in Appendix C) got persistently the reverse order of magnitude of gaps, confusing the selection of materials with target band gaps for tandem solar cells. Likewise, for MAPbI$_3$, calculations of M-cubic structures (using high-precise GW functional with spin-orbit coupling) gives a 530 meV *lower* gap than the calculated gap of orthorhombic structure. Experimentally it is 40 meV *larger* [5]. Table S-3 in Appendix C shows that this inconsistency persists when advanced exchange and correlation functionals are used.

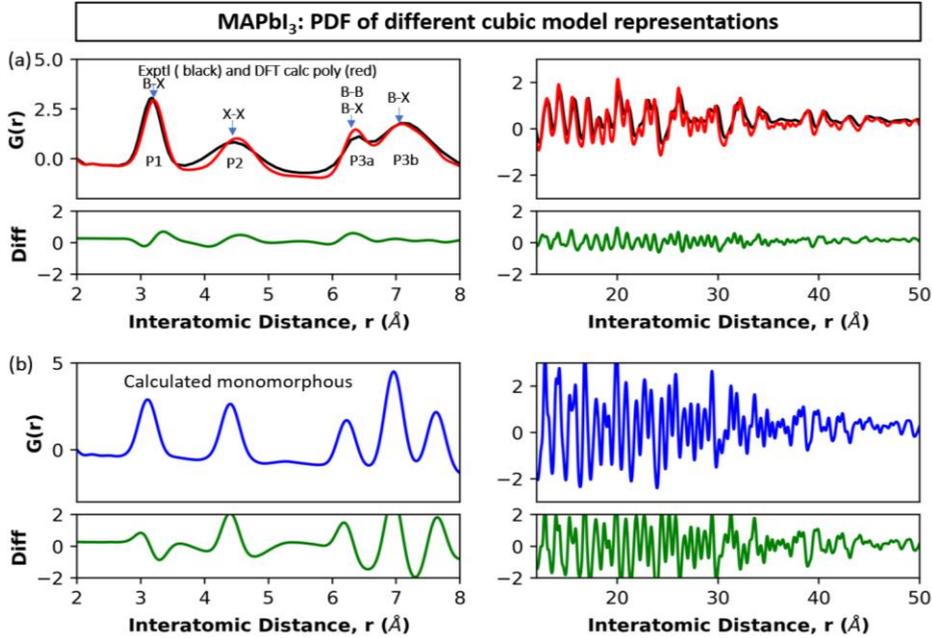

**Figure 2.** Measured [40] (black line in a), and DFT-calculated (a and b) PDF, G(r) for MAPbI$_3$ as function of interatomic distance, r shown at low interatomic distances (left panels) and longer interatomic distances (right panels) with uniform isotropic parameters $U_{ii}$ = 0.01 Å$^2$ for all the atoms. The red line in (a) refers to the DFT predicted PDF of the 32 f.u./cell polymorphous network. As described in Appendix-A material S-1a, 32 additional f.u.("padding") were added non-self consistently to the central cell to reduce periodicity errors. The vertical arrows in (a) indicate bond lengths in the ABX$_3$ structure. (b) is calculated within the monomorphous 1 f.u. cubic cell and deviates significantly from experiment, especially in the 6-8 Å region of the P3a (*i.e.*, Pb-Pb and Pb-I) and P3b (*i.e.*, Pb-I) peaks. The difference between experiment and calculated PDF are depicted using green lines.

**Table I.** Experimental band gaps ($E_g^{exp}$) for cubic phase, and calculated band gaps by using PBE functional for different structures of $CsPbI_3$, $CsSnI_3$, $MAPbI_3$, $FAPbI_3$ and $FASnI_3$. Note that tetragonal and orthorhombic structures are all monomorphous.

|  | $E_g^{exp}$ (eV) | Monomorphous-cubic $E_g$ (eV) | Polymorphous cubic $E_g$ (eV) | Tetragonal $E_g$ (eV) | Orthorhombic $E_g$ (eV) |
|---|---|---|---|---|---|
| $CsPbI_3$ | 1.73[a] | 1.32 | 1.86 | 1.43 | 1.81 |
| $CsSnI_3$ | **1.30[b]** | 0.27 | 0.92 | 0.43 | **0.82** |
| $MAPbI_3$ | 1.57[a] | 1.37 | 1.83 | 1.47 | 1.67 |
| $FAPbI_3$ | 1.48[a] | 1.41 | 1.70 | 1.38 | -- |
| $FASnI_3$ | **1.41[b]** | **0.43** | 1.06 | 0.49 | -- |

a: Snaith, Energy Environ. Sci., 2014,7, 982-988 (ref [47]) b: Kanatzidis, Inorg. Chem. 2013, 52, 9019 (ref [48])

**Table II.** The differences $\Delta E_{tot}$ of total energy and $\Delta E_g$ (meV) of band gap between the large supercells with relaxation (polymorphous network) and the corresponding minimal cell structure (monomorphous approximation) by using PBE functional. The small positive values of $\Delta E_{tot}$ in the third and fourth columns are due to the numerical uncertainties (*e.g.*, k-mesh of different lattice types). The estimated uncertainty is 20-30 meV/f.u. for total energies and ~0.15 eV for band gaps.

|  | $\Delta E_{tot}$ (meV/f.u.) | | | $\Delta E_g$ (meV) | | |
|---|---|---|---|---|---|---|
|  | Cubic | Tetragonal | Orthorhombic | Cubic | Tetragonal | Orthorhombic |
| $MAPbI_3$ | -72 | 2 | 4 | 460 | 30 | 10 |
| $CsPbI_3$ | -123 | 9 | 6 | 540 | 30 | 10 |
| $CsSnI_3$ | -55 | 0 | 14 | 650 | 90 | 80 |
| $FAPbI_3$ | -149 | -1 | -- | 290 | 20 | -- |
| $FASnI_3$ | -144 | -10 | -- | 660 | 70 | -- |

*Greatly overestimated temperature band gap renormalization energy:* When the M-Cubic phase was used to approximate the low temperature phase, the calculated band gap renormalization energy (difference in band gaps at high vs low temperature) turned out to be 390-640 meV [6], far larger than the measured values [5], typically ~50-200 meV.

*Reverse sign of mixing enthalpy:* The alloy mixing enthalpy $\Delta H(A_xB_{1-x})$ measures the enthalpy $H(A_xB_{1-x})$ of an $A_xB_{1-x}$ alloy taken with respect to equivalent amounts of the energies of the constituents $xH(A)+(1-x)H(B)$. When $H(A)$ and $H(B)$ were calculated from the macroscopically averaged M-cubic structure [39], the resulting $\Delta H(A_xB_{1-x})$ is often negative, implying long range ordering at low temperature, which was never observed.

*The macroscopically averaged configuration is phonon unstable:* Some cubic perovskite phases can be stable as an intrinsically monomorphous phase, *e.g.*, $SrTiO_3$ or $BaZrO_3$. That the M-Cubic structure cannot be physically realized in some other perovskites is obvious from the appearance of numerous dynamically unstable phonon branches in a broad range of wavevectors in the DFT calculated harmonic phonon dispersion curves [49] shown in Fig. S-1 in Appendix A S-1b (similarly unstable phonons were reported by refs [34,50–52]). Such dynamic instabilities imply that the cubic Pm-3m monomorphous structure must be replaced by a stable structure, as disscussed for semiconductors in Ref [53]. At low temperature, the replacing structures are the orthorhombic or trigonal ground state phases, whereas at high temperature, anharmonic phonon-phonon interactions can stabilize the cubic $CsSnI_3$ [14].

*The monomorphous structural anzats was used in a wide range of electronic structure calculations*: Despite the inconsistencies surrounding the use of monomorphous structures described above, such structures continue to be widely used as input to electronic band structure calculations and to X-ray refinement. This is in part because one cannot use the structure deduced from PDF fits because these structures are non-periodic. Consequently, the monomorphous representation of single unit cell is very popular for both X-ray [54–56] and neutron diffraction refinement [18], calculation of PDF [19], calculation of band structures [20,21,29,31,33,57,58] and band offsets [24,25], phonons [34,50–52], qualitative description of trends based on crystal symmetry [24], including also topological properties [59], and high throughput calculations and machine learning for discovery and design [23] of new halide perovskites.

We next show how the replacement of the monomorphous ansatz by the lower energy polymorphous network removes these and other inconsistencies.

### III. The energy stabilization of the polymorphous structure

Since the high temperature thermally disordered cubic phase (described, *e.g.*, via molecular dynamics) cannot be properly thought of as developing from the monomorphous, 1 f.u./cell structure that is fictive, we seek to identify the kernel low $T$ structure from which the high $T$ behavior emanates. To do so we search for the structure that minimizes the $T = 0$ *internal* energy of the cubic lattice (Fig. 1d at low $T$), without restricting the cell to a single formula unit, a restriction that does not allow general octahedral deformations without violating periodicity. We thus increase the cell size from 1 f.u. (5 atoms), while retaining the cubic supercell lattice vectors. To avoid locking into a local minimum, we apply random atomic displacements (up to 0.15 Å) before total energy/force minimization is applied. Note that the explored DFT Born-Oppenheimer surface includes, in principle, full anharmonicities. As in any calculation of a supercell with positionally relaxed atoms, here too there are standard convergence tests as a function of supercell size and the largest magnitude of force $F$ allowed on atoms that is used as criterium for declaring the calculation converged. For the latter we use $F_{max} < 0.01$ eV/Å.

In conventional, intrinsically monomorphous compounds such as III-V or II-VI compounds such as GaAs or rock salt MgO or zincblende ZnS, the total energies per formula unit of a large supercell and small supercell are the same, within computational precision. We see from Table II that for halide perovskites, the low temperature ground state structures (orthorhombic, tetragonal) are intrinsically monomorphous, *e.g.*, they do not develop additional local motifs. The total energy per formula unit of these phases does not change within computational uncertainty if calculated in small or large supercells. On the other hand, the cubic phases are not stable in a monomorphous structure. The total energy of the respective *cubic phases* is stabilized by polymorphism by up to $\Delta E_{tot}$ ~140 meV/f.u. Moreover, as illustrated in Appendix-D Fig. S-4, the total energy of the cubic $CsPbI_3$ converges as a function of the supercell size for 32 f.u./cell, (160 atoms) indicating that our polymorphous structure with 32 f.u./cell should be sufficient to capture all possible local deformations. Indeed, the polymorphous state is not an excited state but an alternative to the fictitious cubic monomorphous model.

*Are the properties of a polymorphous network unique for different polymorphous realizations*? In an ensemble configuration such as the polymorphous network, it is possible that the total energy surface does not yield a narrow and sharp minimum. corresponding to a single structure. To investigate the properties of independently generated polymorphous networks we use cubic, 32-f.u./cell (160-atom/cell) $CsPbI_3$ supercells as example: we minimize the internal DFT total energy for different *randomly selected initial nudges* (random on both direction and amplitude) on each atom and then relax all atomic positions (while constraining the macroscopic cubic lattice vectors) until the forces on all atoms within the supercells are below our standard tolerance of 0.01 eV/Å. To check robustness, we start our minimization by independently generated random initial displacements, and relax all of them to the minimum force configurations.

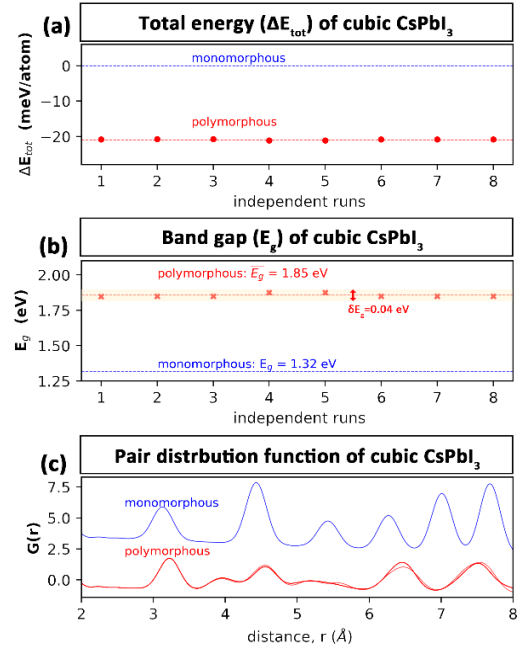

**Figure 3.** The spread of (a), total energy (b) band gap and (c) pair distribution function (PDF) for the cubic $CsPbI_3$ (32 f.u./cell) for independently generated polymorphous networks using different initial random nudges with randomly selected orientations and amplitudes ranging from 0.00 to 0.15 Å in steps of 0.01 Å. Results are compared to the monomorphous structure. The PDF is calculated using a uniform isotropic parameters $U_{ii}$=0.01 Å$^2$).

To test the possible spread in properties, we apply to each polymorphous supercell atomic relaxations by first nudging each of its atoms off site along a randomly selected direction by amplitudes between 0 to 0.15 A, followed by complete cell internal relaxation (to force $F_{max} < 0.01$ eV/Å), constraining the supercell to its macroscopically observed cubic shape. To find if different relaxations produce the same (or different) polymorphous displacements, we restart 8 independent calculations, each with its different, randomly selected and nudged starting configurations, which is then relaxed to completion. In some cases, repeating the nudge is needed to achieve minimum force. Fig. 3a shows the total energy spread, figure 3b shows the spread in band gap, and Fig. 3c shows the spread in calculated PDF. We see that the total energy variations among different randomly selected polymorphous realizations is < 0.1 meV/atom, with average deviation of 0.05 meV/atom, a difference that lies within the calculation error tolerance, and is much smaller than the total energy difference between the monomorphous structure and the (average) polymorphous structure.

(we note that using a large amplitude nudge, as we initially tried, can converge much slower, posing difficulties to reach a unique configuration). Likewise, the band gaps of the different polymorphous realizations have a spread of 0.04 eV. Again, this spread is negligible relative to the band gap difference between the monomorphous structure (1.32eV) and the (average) polymorphous structure (1.85eV). The different polymorphous realizations give consistent PDF in the distance range of 3.5-8.0 Å, and the calculated PDF is very different than PDF generated from the monomorphous structure. We conclude that the static relaxation and the physical properties calculated from different polymorphous realizations are robust.

### IV. The PDF of the energy-minimizing polymorphous network

To examine if the energy minimizing geometry is realistic, we compare in Fig. 2 the ab-initio calculated pair distribution function (PDF) with the experimental result [38] (shown by the black line in Fig. 2a), both for short and intermediate interatomic separations (left side panels) and long range separations (right side panels).

Previous PDF measurements and simulations [60] often fit the data in a small unit cell by introducing parameters [40] such as atoms having non spherical shapes, even changing the values of the observed lattice parameters, and altering Wyckoff positions with respect to XRD measurement in search for an effective model that fits the PDF. The ensuing structural models are generally non-periodic, and the description of atoms as finite, shaped objects is incompatible with the way crystal structures are used as input in periodic band structure calculations. Thus, no bridge connects such PDF to other predicted properties. The present approach does not face this difficulty as we use a fully constrained energy minimizing approach with respect to all 3N-6 degrees of freedom for N atoms without introducing any fitting parameter. To minimize finite size effects, we use a padding of the relaxation-active central supercell by additional bulk-like halide perovskite material all around the central cell. This reduces periodicity errors.

We recall that sometimes the procedure of generating a crystal structure from PDF measurements used *large cells* as well[29] but this was a fit to experiment, not a prediction. Such fitting procedures with large cells face the problem of over fitting because of insufficient data with respect to the number of fitting degrees of freedom. This problem does not exist in variational total energy minimization that uses large supercell with hundreds of atoms (≥160 atoms/cell), offering the opportunity to stick to the conventional description of atoms as shapeless points in a periodic array; the ensuing structure is obtained by first principles, not fitting, and then directly useable in electronic structure calculations using any standard band structure method. This is consequential as it provides a direct bridge between PDF and electronic properties.

We see from Fig. 2b, that the monomorphous structure (with uniform isotropic parameters, $U_{ii} = 0.01$ Å$^2$ for each Pb and I atom to roughly include the finite temperature effect) gives rather poor agreement with experiment (see, in particular the split three peaks at 6-8 Å). The red line in Fig. 2a shows the PDF calculated from the polymorphous network with the same isotropic parameters. Given that our large supercell calculation is generated from total energy minimization, we consider the agreement in Fig. 2a between theory and experiment as good.

The crystallographic structure file that produces this good agreement with the observed PDF has been used for all calculations of the electronic structure in this paper.

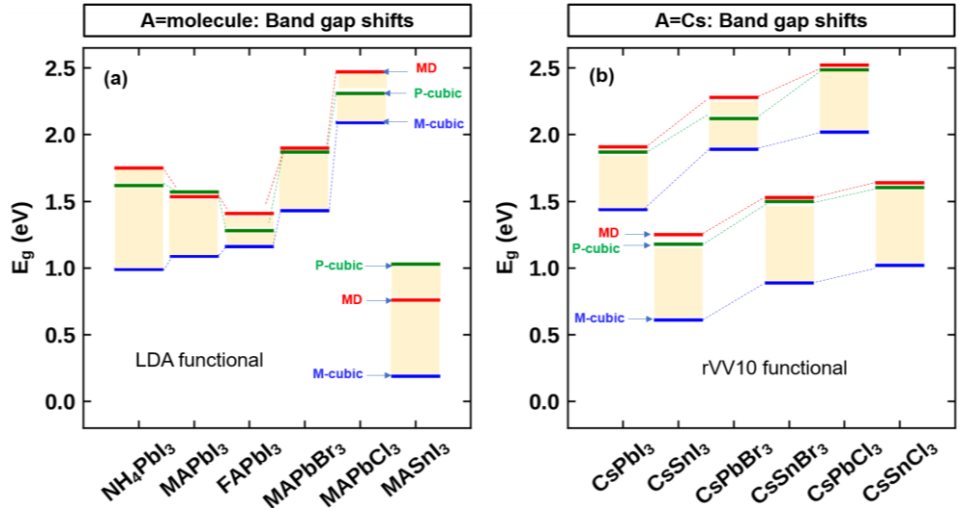

**Figure 4.** Band gap values of the DFT monomorphous cubic (M-cubic, blue), the polymorphous cubic (P-Cubic, green), and literature Molecular Dynamics (MD, red). The MD values in (a) come from Mladenović *et al*. [33] who used the LDA functional); and in (b) from Wiktor *et al*. [6] who used the rVV10 functional. NH$_4^+$ is included as an example of a molecule with small effective radius even though it does not have the perovskite structure. All values for M-cubic and P-cubic were calculated with the functionals as in the respective MD results.

## V. Consequences of cubic polymorphous networks on electronic properties

Because the polymorphous networks manifest lower local symmetries than the global averaged symmetry sensed by X-Ray diffraction (embodied by the fictitious monomorphous structure), polymorphous electronic structure calculations that 'see' local symmetries produce new, previously unappreciated features:

*Correcting the band gap anomaly:* Polymorphism significantly increases the band gap relative to the monomorphous ansatz (Fig. 4). We note that DFT does not produce accurate absolute band gaps, in particular when spin-orbit coupling is neglected. Here we focus on the change in band gap due to allowing a polymorphous network. Here we do not focus on getting the exact absolute magnitudes of the band gaps but wish instead to understand the importance of the structural representation on the trends in the gaps in a series of compounds (as shown in Table I and Appendix-C Fig. S-3). We find that, when the compound is intrinsically polymorphous (as in the cubic phase of the perovskites), using the incorrect monomorphous structure as input, one can underestimated the gap by up to 300%. To qualitatively assess which compounds have the largest increase in band gap in polymorphous relative to monomorphous, Appendix-E figure S-5 depicts the respective band gap shifts $E_g$(Poly) - $E_g$(Mono) as well as $E_g$(MD) - $E_g$(mono) *vs* the tolerance factor [61]. This shows that compounds that have tolerance factor significantly lower than 1.0 (*i.e.*, likely unstable in the monomorphous structure) have the largest polymorphic effects, consistent with the view that polymorphism is associated with cubically unstable structures.

*Reducing significantly the band gap renormalization error:* The shifts of band gaps due to temperature are generally calculated as the shift between finite temperature gap from molecular dynamics ($E_{MD}$(T), *i.e.*, the averaged band gap value of the equilibrium configurations appeared in the MD trajectory), and a low temperature reference band gap ($E_{LT}$), which is usually attributed to the thermal effect. Fig. 4 also shows the comparison of the band gaps obtained from the polymorphous network (T=0, minimization of the internal energy alone), with the band gaps obtained in the literature using finite temperature MD [6,42]. We used precisely the same exchange correlation functionals and lattice constants as in the respective MD calculations to assure consistency of the results. We see that the MD gaps are slightly higher but very similar to the P-Cubic gaps, consistent with the view that the polymorphous structure derived from minimizing the *internal energy* captures the leading spectrum of distortions that control the band gaps at higher temperatures. Additional thermal disorder effects introduced specifically by entropy lead to additional small increase in the band gap as temperature rises. In contrast to what Mladenović et al. [42] and Wiktor [6] et al. did (use a monomorphous structure at *T*=0 as reference to calculate the renormalization energy with respect to the MD gap at high *T*), we define the renormalization energy as the difference of MD gap with respect to the polymorphous network. The latter approach we use gives ~200 meV renormalization, very close to experiment, while the approach of Wiktor et al. [6] gives a 390-640 meV renormalization.

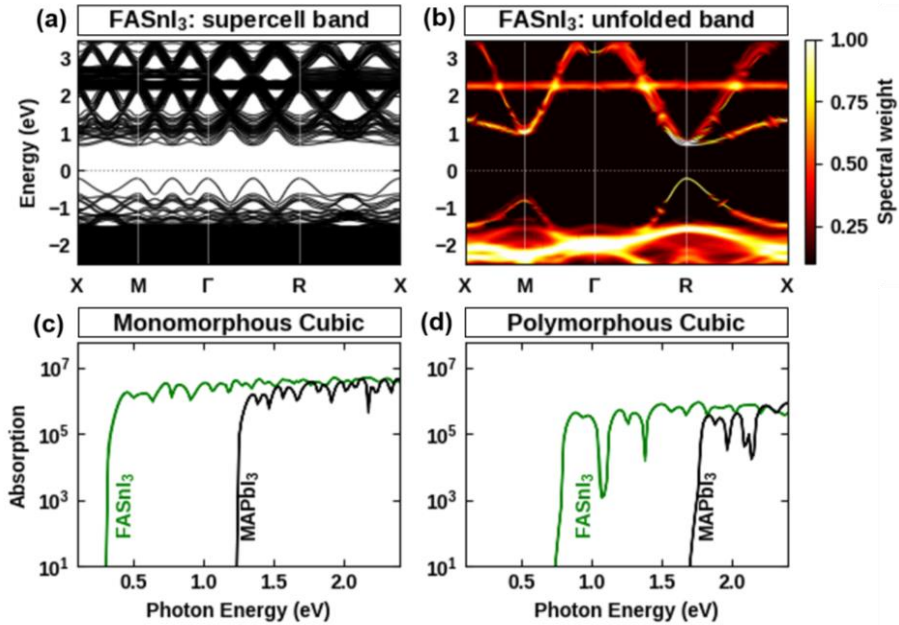

**Figure 5.** (a) The 'Spaghetti-like' band structure of FASnI$_3$ supercell (32 f.u./cell). (b) The effective band structure (EBS) of the same FASnI$_3$ supercell, unfolded to the primitive Brillouin zone. The absorption spectra of (c) monomorphous cubic (1 f.u./cell) and (d) polymorphous cubic (32 f.u./cell) FASnI$_3$ and MAPbI$_3$.

*Restoring the correct trend among the band gaps of FASnI$_3$ vs CsSnI$_3$*: The anomalous orders of band gaps of cubic FASnI$_3$ *vs* orthorhombic CsSnI$_3$, and that of cubic MAPbI$_3$ vs orthorhombic MAPbI$_3$ result from modeling the cubic phase in a monomorphous fashion. This is fixed by using the correct P-cubic gaps (Fig. S-3 in Appendix-C and Table I) showing that, in agreement with experiment [5,62], the band gap of CsSnI$_3$ is smaller than that of FASnI$_3$, whereas the band gap of CsPbI$_3$ is larger than that of FAPbI$_3$. These opposing trends can be understood by noting that octahedral distortions raise the band gap, and by analyzing the relative distortions for each of these compounds, shown in Fig. 1 for FASnI$_3$ and for other halide perovskites in Appendix-F Fig. S-6. The distortions calculated for FASnI$_3$ are much larger than those of CsSnI$_3$ (mainly off-center), leading to a larger band gap in the former. On the other hand, the trend is the opposite for the Pb compounds, having larger tilting angles in CsPbI$_3$ compound relative to FAPbI$_3$ and consequently the former presenting larger energy gaps.

*Band edge states of the polymorphous network remain sharp, as does the absorption onset:* An attractive feature of the monomorphous cubic (Pm-3m) structure is that its absorption spectrum [63,64] shows a sharp rise, which has been long used to explain why these materials are good solar energy absorbers. One might expect that the polymorphous 'structural disorder' associated with different distortions for different octahedra would create localized gap states as in traditional disorder models [65,66]. This is difficult to evaluate from a supercell band structure (Fig. 5a) that folds-in a large number of bands, making it impossible to visually extract the *E vs k* dispersion.

To examine the important *E vs k* dispersion, we consider the Effective Band Structure (EBS) concept of band unfolding [67,68], yielding the EBS in the primitive Brillouin zone. One unfolds the supercell band structure by calculating the spectral weight $P_{Km}(k_i)$ and from them the spectral function $A(k_i, E)$, as illustrated in Appendix-A S-1c. Like the experimental Angle-Resolved Photoemission Spectroscopy, EBS provides a 3-dimensional picture of spectral weight with coherent and incoherent features, all naturally produced by the polymorphous network consisting of the many local environments.

Fig. 5b shows the EBS of a 32 f.u./cell of FASnI$_3$ folded into the primitive Brillouin zone of the single cell. Every band now shows a mixed characteristic of coherent, 'sharp' dispersion and incoherent, 'fuzzy' broadening. *Remarkably, the band edges (R point) are sharply dispersiv*e despite the existence of a distribution of deformations that might appear as 'disorder', indicating the small Urbach tails that is consistent with experimental observation and molecular dynamic simulation. We suggest that it is due to antibonding states at VBM and CBM of halide perovskites, in which cannot easily form in-gap states that contribute to Urbach tails. Concomitantly, the calculated absorption spectra in Fig. 5c and 5d (details in Appendix-A S-1d) of FASnI$_3$ and MAPbI$_3$ in the polymorphous cubic structures show sharp absorption edges, similar to that of the (fictive) undistorted monomorphous structure. This is in accord with the performance of such halide perovskites as superb absorbers [66], and in contradiction with the naïve view that octahedral deformations are a form of disorder.

Two factors may contribute to this. First, octahedral deformations *increase* the band gap by moving the VBM to deeper energies (Fig. S-7), so this particular type of structural displacements shift the would-be localized states into the band continuum, not the gap region. Second, the octahedral distortions couple only weakly to the band edge wavefunctions at R point. This is suggested by the great similarity in the hybridization of band edges (B-*s* with X-*p* in the VBM, and B-*p* with X-*p* in the CBM) before and after distortions, as shown in Appendix-G Table S-7 and Appendix-H Fig. S-8 presenting large joint density of states and strong transition amplitude. This suggests that polymorphous octahedral distortions do not act as "conventional disorder": while they shift band edges, they do not create localized gap states or band tail.

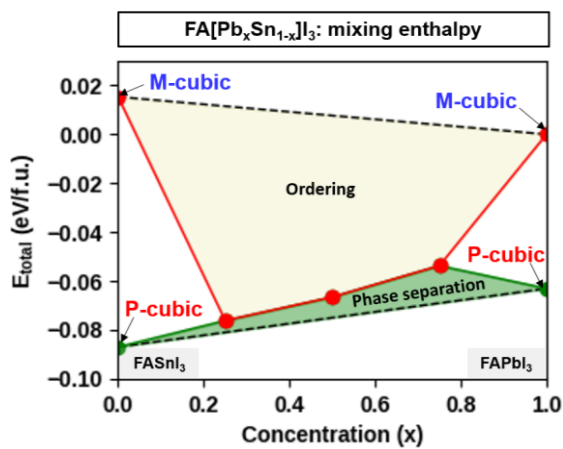

**Figure 6.** Alloy mixing enthalpies of FASnI$_3$ and FAPbI$_3$, with respect to monomorphous cubic (M-cubic) phases (yellow region) and polymorphous cubic (P-cubic) (green region), showing that the M-cubic values for the pure, non-alloyed compounds leads to negative enthalpy (implying ordering at low temperature), whereas the polymorphous network correctly predicts small positive mixing enthalpies (implying phase separation).

*Reconciling the contradiction about Phase-separation vs ordering:* Another physical property that depends on the definition of the structure of ABX$_3$ is the alloy mixing enthalpy $\Delta H(A_x B_{1-x})$ that measures the enthalpy $H(A_x B_{1-x})$ of an $A_x B_{1-x}$ alloy taken with respect to equivalent amounts of the energies of the constituents $xH(A) + (1-x)H(B)$. When $H(A)$ and $H(B)$ are calculated from the high-energy macroscopically averaged M-cubic structure [39,69–71] the resulting $\Delta H(A_x B_{1-x})$ was often negative (Fig. 6, yellow shading) implying long range ordering at low temperature, which was never observed in halide perovskite alloys (either A-site, or B-site, or X-site alloys). Instead, phase segregation has been measured in experiment (*e.g.*, for X site mixed alloys [70]). Indeed, calculated excess enthalpy of mixed anion alloys MAPb($I_{1-x}Br_x$)$_3$ based on the monomorphous assumption by Goyal et al. reported negative alloy excess enthalpy [39], implying tendency of long-range order, which was never observed; instead, phase segregation has been measured for X site mixed alloys [70]. As to mixed B-site alloys such MAPb$_x$Sn$_{1-x}$I$_3$, no long-range order was observed and the existing samples of random alloys appear as single-phase disordered that are entropy stabilized [69]. Because of the availability of this data, we calculated the same B site alloys FAPb$_x$Sn$_{1-x}$I$_3$ in cubic phase finding that using the fictitious monomorphous structure (red lines and yellow shading) gives negative excess enthalpy, signaling incorrectly long rang order, but using the polymorphous structure (green lines and green shading) gives (small) positive excess enthalpy [72], suggesting entropy stabilization at intermediate temperatures and phase separation at low temperatures (the latter may be difficult to observe because of the low atomic mobility at such low temperature). Thus, the monomorphous structures yield a contradiction with observed phase behavior, whereas the polymorphous is consistent with available data.

***Enhancement of the dielectric constant and increased role of ionic vs electronic contribution in CsSnI$_3$:*** The near ferroelectric halide perovskites have large static dielectric constants whose magnitude is an important quantity in optical and defect theories. Our calculation of CsSnI$_3$ (Table III) shows that static dielectric constants of the assumed monomorphous phase is rather small ~35, with a 7:10 ratio between ionic and electronic contributions.

However, for the polymorphous network we find a much higher static dielectric constant of 128, with a ratio of 16:1 between ionic and electronic contributions. This is close to the low frequency measurements of MAPbI$_3$ at high temperatures [73]. Given hydrogen-like Wannier-Mott exciton model [74], large static dielectric constant [73] leads to small exciton binding energy, which benefits the separation of electron and hole.

**Table III.** Calculated averaged macroscopic dielectric constants originated from ionic and electronic contributions of CsSnI$_3$ by using DFPT theory for different phase of CsSnI$_3$.

| CsSnI$_3$ structure | ε(ionic) | ε(electronic) |
|---|---|---|
| Monomorphous (1 f.u./cell) | 35.11 | 54.97 |
| Tetragonal (Monomorphous) (2 f.u./cell) | 32.78 | 15.19 |
| Orthorhombic (Monomorphous) (4 f.u./cell) | 28.14 | 7.61 |
| Polymorphous (32 f.u./cell) | 128.23 | 8.08 |

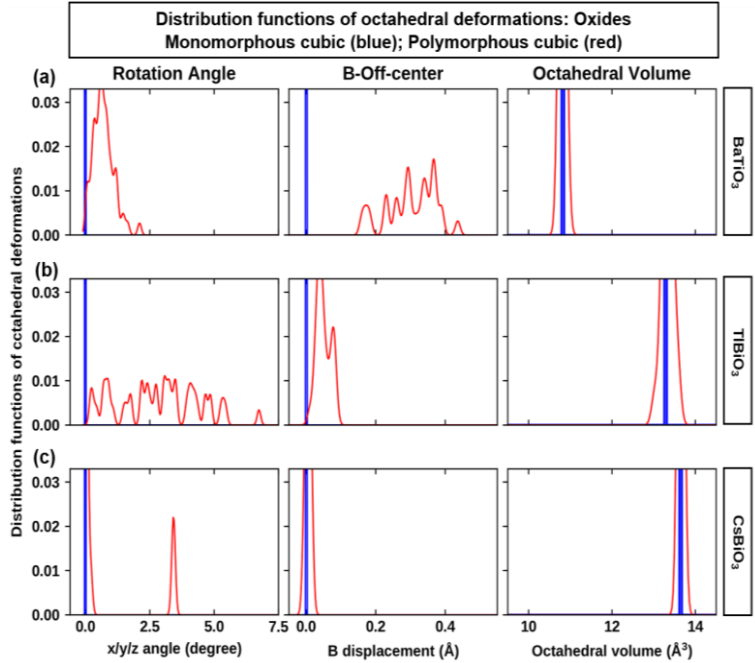

**Figure 7.** Statistics of motifs in perovskites for (a) BaTiO$_3$ (t=0.94), (b) TlBiO$_3$ (t=0.95) and (c) CsBiO$_3$ (t=1.01).

## VI. Not only in halide perovskites

The discussion on polymorphous networks in cubic halide perovskites focuses on *positional* polymorphism. There is also spin polymorphism (ie different local spin environments) noted earlier for electronic spin in paramagnetic $3d$ oxide [75,76] and in the paraelectric electronic polarization [77,78]. The polymorphous cubic phases apply also to oxides that have dynamically unstable phonons, as shown in Fig. 7(a, b), *i.e.*, $BaTiO_3$, $TlBiO_3$, showing that this effect is not limited to halides. For comparison, we calculated also the relaxed total energy and band gaps of a compound that has *stable phonons* in the monomorphous structure, *i.e.*, $CsBiO_3$ [49]. $BaTiO_3$ was known [79] to manifest tilting and rotation, and our energy minimization of Fig. 7 shows that this is accommodated by developing a range of local environments, in particular for B site displacements. $CsBiO_3$ and $TlBiO_3$ were inferred to exist by Machine learning [49] but are not otherwise known compounds. We add them as calculated predictions for polymorphous behavior, correlating this with phonon stability *vs* instability in the hypothetical monomorphous structure. $CsBiO_3$ has stable phonons in the monomorphous structure and Fig. 7c shows that there are no polymorphous characters, *i.e.*, there is but uniformed tilting angle, negligible B-site off-center and change of volume. $TlBiO_3$ has phonon instability in the monomorphous structure and shows in Fig. 7b characteristic polymorphous behavior.

We note that the related subject of nematicity—lowering of symmetry observed through electronic structure (local) probes with respect to the perceived global symmetry- as noted for example in FeSe [80,81] is potentially closely related to polymorphous networks, where the existence of different local environments can lead to electronic symmetry breaking despite the existence of higher *average* symmetry noted by global probes. Because polymorphism removes the centrosymmetric symmetry, one would expect also a Rashba effect and related optical measures of lowered symmetry.

## VII. Discussion of the contribution of intrinsic a-thermal displacements *vs* thermal disorder

This paper deals with a prototypical system with two phases, (a) the low temperature ground state (GS) structure (here, orthorhombic or tetragonal for perovskites) that is a fully ordered structure at $T=0$ K, and (b) the high temperature cubic phase. We are interested here primarily in the latter cubic phase, that is of general interest as a light-absorbing phase and shows a polymorphous network, whereas the low temperature ground state structure does not (see Table II showing no total energy lowering for the ground state as the cell increases). Each of the two phases is discussed in terms of by (i) its internal energy $H$ that has but a small temperature-dependence that will be neglected here, and (ii) its free energy $F = H - TS$ that has an additional entropy-component $-TS$ of disorder when $T > 0K$ that is commonly described by Molecular Dynamics (MD) simulations [7–13].

*The physics emerging from minimizing H:* We first evaluate the properties of each of the two phases by constraining the unit cell shape to the phase of interest (tetragonal, cubic, others), and minimizing its respective $H$, described by DFT, as a function the cell-internal atomic positions. This provides the ideal low temperature atomic arrangements of the respective phase, *i.e.* the structural preference dictated by the specific chemical bonding in the system, before the effect of thermal disorder sets it.

A central observation is that allowing a large $N$ x $N$ x $N$ 'supercell', leads upon minimization of $H_{cubic}(T=0K)$ to a considerably lower energy (the red line in Fig. 1d denoted $H^{(P)}_{cubic}$) than what is obtained by minimizing a minimal unit cell (1 f.u./cell), shown in Fig. 1d by the blue line denoted $H^{(M)}_{cubic}$. This energy lowering saturates for large enough supercell ($N\sim32$ formula units) and distinguishes these compounds from the vast majority of inorganic crystals where a primitive unit cell description captures the real structure. The lowering of enthalpy of the cubic phase afforded by the supercell representation reflects the intrinsic chemical bonding preference—here, the B atom lone pair that tends to create different local bonding environments. This breaking of the local symmetry is manifested by a set of octahedral tilting, rotations, and B atom displacements and is enabled by the supercell representation that allows the needed spatial flexibility lacking in the traditionally restricted minimal unit cell. We refer to this structure type as *polymorphous* ('many forms') *network*, whereas the artificial, high energy, high symmetry "virtual crystal" structure that is restricted by symmetry to have a single repeated motif with energy $H^{(M)}_{cubic}$ will be referred to as monomorphous. Large unit cells with many Wyckoff positions are not a stranger to crystal chemistry [82], even though often X-ray diffraction, in taking a spatial average of significant sample volume, is fit to a deceivingly high-symmetry monomorphous structure. This pattern of displacements has a different origin than the dynamic motion due to temperature (see discussion of $G=H-TS$ below) that yields displacements about the ideal structural position, and averages over snapshots to the ideal, non-displaced structure. The ground state structure of halide perovskites (here, orthorhombic) does not lower its internal energy $H_{GS}$ by increasing its nominal cell size, i.e. this structure in inherently monomorphous.

The present paper studies the properties of the atomic position polymorphous network of a few cubic halide and oxide perovskites, including optical, dielectric, structural and thermodynamic properties, finding that, relative to the virtual monomorphous structure, widely used in the literature a broad range of electronic structure calculations, use of the former removes many outstanding conflicts with experiment. This establishes such cubic phase of these compounds as being inherently *atomic-displacement polymorphous*, due to the nature of its chemical bonding. This draws an analogy with the recently discovered *spin configuration polymorphous networks*, characterizing paramagnetic $ABO_3$ "Mott insulators" with different local *spin* environments [75,76]. Indeed, the polymorphous network is not a model for approximate description but a physical structure. Yet a useful point is that the minimization of the internal energy under the constrain of macroscopically cubic phase provides a very useful approximate to the physical configuration. This can be used to calculate, with standard DFT codes, the structural properties (PDF, Fig. 2), band structure (Fig. 5), alloy physics (Fig. 6), band gaps (Table I), dielectric constant (Table III) in substantial agreement with experiment, and to remove many of the inconsistencies that existed previously between monomorphous DFT calculations and experiment. Thus, the polymorphous approximate could serve as a very useful practical structure to use with standard band structure approaches to predict properties. Thermal agitation is an additional contribution to our a-thermal descriptor (Fig. 4).

***The physics emerging from H-TS:*** The ordered pattern of displacements that emerge from minimizing $H_{cubic}(T=0K)$ is the kernel of the structure that evolves from it thermally via stochastic dynamic displacements (e.g., seen in MD). The time- and spatial-average of the dynamic motions would yield the starting static structure, whereas the static structure obtained by minimizing $H_{cubic}(T=0K)$ does not average to an ideal monomorphous network. Thus, the local motifs seen in the polymorphous network present static distributed deformations, rather than the dynamic/vibrating due to temperature [82].

As temperature is increased, the displacements of the static polymorphous network develop dynamic components. Those were initially thought to be large, because the reference point used to compare MD results was the fictitious monomorphous structure rather than the property minimized polymorphous network. Indeed, discussion of MD results have often obscured the fact that the basic structure from which temperature induced displacements develop is not the traditional monomorphous structure, but the polymorphous one introduced here. We compare materials properties such band structure and band gap as described by the polymorphous structures that minimizes $H^{(P)}_{cubic}(T=0K)$ to the properties obtained in the literature from finite temperature MD (compare red and green lines in Fig. 4), finding that the additional thermal displacements often modify the materials properties rather modestly, suggesting that the essential physics was already established by the internal energy dictated structures[94].

Even though MD simulations could, in principle, mimic the average cubic monomorphous structure at high *T*, the vast majority of these studies [5–13] do not focus on this kind of analysis. Several works focused on the motions of the organic molecules, including the description of 'pair modes' related to their relative alignment [83]. Quarti et al., calculated the average structure of a MD trajectory, but the resulting phase was not monomorphous cubic [5]. As far as we are aware, there are no MD simulations comparing the monomorphous cubic structure to the distorted polymorphous network.

We conclude that using polymorphous networks to replace fictitious monomorphous structures defines a broader principle of the need to describe broken-symmetry systems manifesting a range of microscopic configurations whose physical properties can be very different than the often-assumed macroscopic average.

**Acknowledgement**

The work at the University of Colorado at Boulder was supported by the US Department of Energy, Office of Science, Basic Energy Sciences, Materials Sciences and Engineering Division, under Grant No. DE-SC0010467 to the University of Colorado. Work on photovoltaic relevant absorption characteristics was supported by the U.S. Department of Energy, Energy Efficiency and Renewable Energy, under the SunShot "Small Innovative Programs in Solar (SIPS)" Project No. DE-EE0007366. G.M.D. also acknowledges financial support from the Brazilian agencies FAPESP and CNPq.

**Appendix A: Methods**

**1. Computational details**

All the calculations were carried out by using the projector augmented wave (PAW) [84] method with the Perdew-Burke-Ernzerhof (PBE) [85] generalized gradient approximation (GGA) as implemented in the Vienna ab initio simulation package (VASP) [86,87]. The cutoff energy was set to 520 eV. Energy convergence criterion was set to $10^{-6}$ eV per unit cell, and the forces on all relaxed atoms are less than 0.01 eV/Å. Depending on the lattice parameters, a different k-point mesh was set. To get accurate atomic positions, the van der Waals interaction was considered by using optB86b-vdw functional [88].

For the minimal cubic phases (1f.u./cell) of $ABX_3$ (A=Cs, FA, MA; B=Sn, Pb; X=I) we relax the atomic positions by keeping cubic lattice vectors and symmetry with 6x6x6 k-mesh (lattice constants reported in our previous work [38] were used, and the dipole of the molecules was set along <111> direction). For the minimal tetragonal and orthorhombic phases (4 f.u./cell), we optimized the atomic positions by keeping the symmetry (dipole of the molecules along <100> and <001> directions) with 4x4x2 k-mesh.

The initial supercell structures (32 f.u./cell) were constructed by using a 2√2x2√2x4 supercell of optimized minimal cubic phases, and by 2x2x2 supercell of optimized minimal tetragonal and orthorhombic structures. Then, we randomly nudged each atom (with a displacement smaller than 0.20Å) along x, y, z directions (except organic motifs) on the initial supercell structures. By using these structures as input, we optimized the structures to their minimal energies by using Γ-only k-point and by keeping their lattice parameters fixed. The total energies and bandgaps of these supercell structures with 32 f.u./cell were obtained by using 2x2x1 k-mesh in the Brillouin zone.

In cubic halide perovskites $ABX_3$, the molecular orientation in A site is suggested to be randomly oriented. For the structure with molecules, taking

FASnI$_3$ as example, we used the special quasirandom structures (SQS) method [89] to generate the initial orientation of the dipole of FA molecules in the FASnI$_3$ supercell. The initial dipoles were along the <111> or <-1-1-1> directions. SQS is designed to find a single realization in given supercell size to best reproduce the properties in infinite alloy. As the pair, 3-body, 4-body, *etc.* correlation functions can all be calculated precisely in the perfectly random, infinite alloy, SQS then searches all possible configurations in the *N*-atom supercell to find the best correlation functions compared to the ones in infinite alloy. Therefore, a property *P* calculated from an SQS is not simply a single 'snapshot' but approximates the ensemble average <*P*> from many random configurations. Description and discussion of SQS can be found in ref [89], furthermore, showed that large size SQS gives more reliable results than calculating ensemble averages directly from many small random supercells, because some intermediate range interactions (*e.g.* long-range pairs) in large supercells do not exist in small ones due to size limitation.

We also considered the effect of the organic molecule orientation at A site: we take FASnI$_3$ as example to set up aligned molecular orientation to check the change of total energy and band gap. Compared to the monomorphous phase (repeated 1f.u./cell) without local distortion, the different orientated molecules indeed result in the increase of band gap and decrease of total energy. The difference of band gap for structures with different molecular orientation is ~0.2 eV, close to the fluctuation of energy gap found by using MD and experimental observations [5]. But the gap difference due to molecular orientation is rather smaller than the difference between polymorphous and monomorphous structure. Notably, the polymorphous structure has the largest band gap increment and total energy decrease.

## 2. Pair distribution function (PDF)

To get the calculated pair distribution functions (PDF) based on the crystalline structure, we used the PDFgui software [90]. The experimental resolution parameters $Q_{damp}$ and $Q_{broad}$ were set to 0.0434148 and 0.0164506 following Ref [40], were determined through refinements of the PDF of the nickel standard (i.e., Nickel single crystal). The $\delta_1$, $S_{ratio}$, and $r_{cut}$ were set to 0.0, 1.0 and 0.0. The isotropic parameters ($U_{ii}$) $U_{11}$ = $U_{22}$ = $U_{33}$ = 0.01 A$^2$ were applied on each atom, which is used to describe the atomic displacements due to thermal effect. The correlated motion corrections were not applied in our calculation of the PDF. This is firstly because our calculations agree well with experiment (Figure 2 and Figure S-0) suggesting that it includes the pertinent physics. Secondly, applying the correlation correction requires empirical parameters, but our calculation is parameter-free. Thirdly, the reference PDF fitted calculation of Ref [40], which we use as benchmark showing what can be accomplished if lots of fitting parameters are used, apparently also did not applied this specific correction either, as evidenced by repeating the small box fit of Ref [40] and reproducing their result without adding the correlation correction.

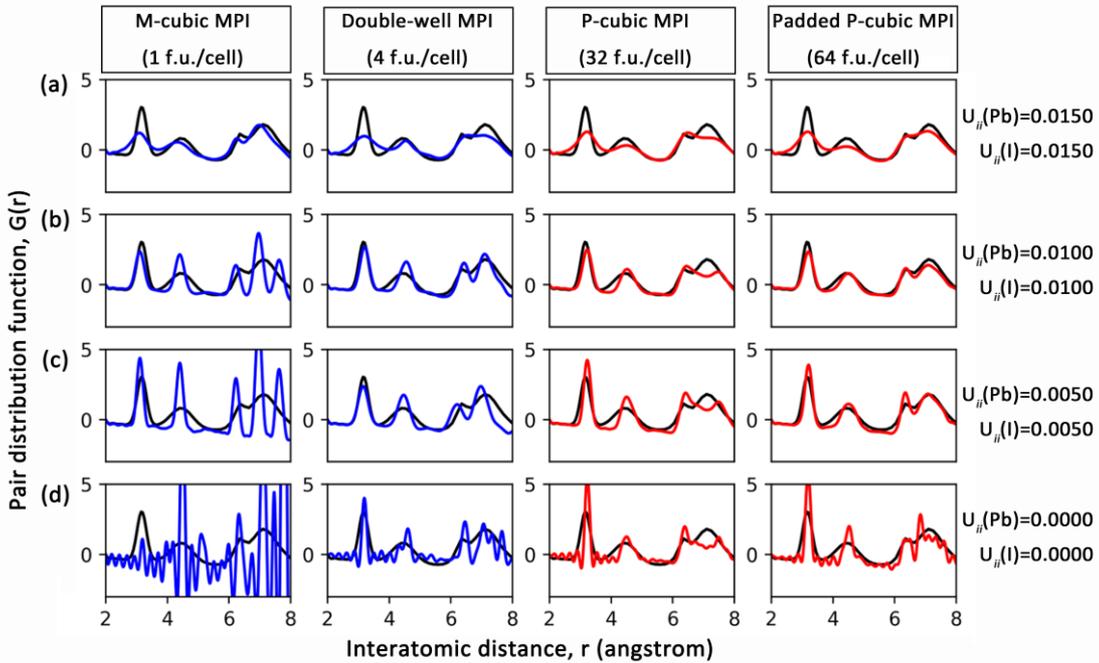

**Figure S-0.** The calculated pair distribution function G(r) as function of interatomic distance (r) with different (plots a-d) uniform isotropic parameters ($U_{ii}$) for monomorphous cubic (M-Cubic; the first column, blue lines), double well (the second column, blue lines) and polymorphous cubic structures (P-cubic) & padded P-cubic structure (the third and fourth columns, red lines). Experimental values are taken Ref [40]. Experimental PDF, G(r) for MAPbI$_3$ (MPI) as function of interatomic distance (r) is represented by black lines.

A large padded structure of MAPbI$_3$, containing 64 f.u./cell, was constructed to overcome the periodicity errors. The padded cubic structure of MAPbI$_3$ includes two parts (Figure S-3): core shell (black) and padding shell (blue). The core is the polymorphous structure (32 f.u./cell), in which all the octahedra/molecules are randomly distorted (tilting, B-site displacement and octahedral volume)/orientation. Padding shell is the monomorphous structure with 32 f.u./cell, in which the octahedra show same tilting and molecules are aligned in the <111> direction. We obtained this configuration by making a 4x4x4 supercell (64 f.u.) with respect to the optimized minimal cell (1 f.u./cell), and then carving a hole (whose size equal to the core-shell configuration) by removing 32 f.u. Then we combined the core and padding-shell, and removed all the molecules. No further relaxation was done for the padded large cubic structure. By using this padded structure and $Q_{damp}$=0.0434148 and $Q_{broad}$=0.0164506 as input, we got the PDF by using PDFgui software [90].

Fig. S-0 shows the calculated PDF of cubic MAPbI3 halide perovskite in different approximations using different isotropic parameters ($U_{ij}$). One can see that as the isotropic parameters, $U_{ii}$ decrease, the calculated PDF are becoming noisier. Generally, isotropic parameters can be calculated from the phonon spectra but more often they are used as fitting parameters. We obtained good agreement on PDF (heights and broadening width of peaks) by using moderately small isotropic parameters for all the atoms based on the padded polymorphous cubic MAPbI$_3$. Note that with different isotropic parameters (or even without isotropic parameters) the polymorphous cubic structure always shows by far the best agreement with experiment, being much better than the double well model or the monomorphous structure.

### 3. Phonon spectra

The phonon spectra for the minimal cubic structures were obtained by using the force constants with density functional perturbation theory (DFPT) and supercell finite displacement method implemented in the Phonopy code. [91] To get the accurate force matrix, a 8x8x8 k-mesh was used for optimizing the minimal structures (1 f.u./cell) with an energy cutoff of 520 eV for the plane-wave basis and an energy convergence criterion of $10^{-8}$ eV considering the van der Waals interaction by using optB86-vdw correction. The force threshold was set to $10^{-4}$ eV/Å. To get the converged force constants, we have performed the phonon dispersion calculations by using a 4x4x4 supercell with 2x2x2 k-mesh, which is sufficient to take into account the long-range interaction.

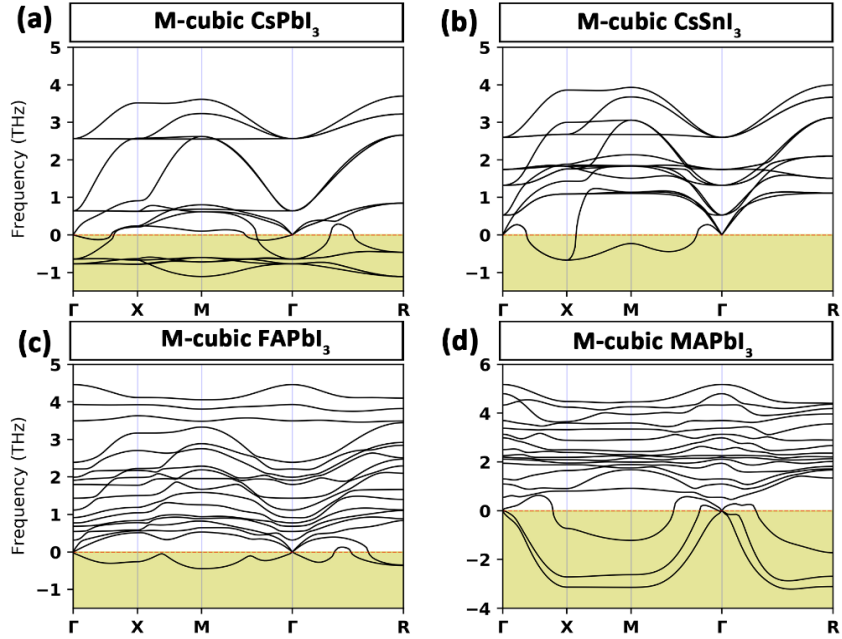

**Figure S-1.** DFT calculated harmonic phonon dispersion curves for monomorphous cubic (M-cubic) halide perovskites (a) CsPbI$_3$, (b) CsSnI$_3$ (c) FAPbI$_3$ and (d) MAPbI$_3$. The shadow region refers to the imaginary frequencies.

### 4. Effective band structure (EBS) [67,68]

The band folding mechanism in a supercell can be expressed as

$$|Km\rangle = \sum_{i=1}^{N_K} \sum_n F(k_i,n;K,m)|k_i n\rangle \quad (1)$$

where $|Km\rangle$ is the $m$-th electronic state at $K$ in supercell Brillouin zone, $|k_i n\rangle$ is the $n$-th electronic state at $k_i$ in the primitive Brillouin zone. One can then unfold the supercell band structure by calculating the spectral weight $P_{Km}(k_i)$ from

$$P_{Km}(k_i) = \sum_n |\langle Km | k_i n \rangle|^2 \qquad (2)$$

which is the Bloch 'preservation' of Bloch wavevector $k_i$ in $|Km\rangle$ when $E_n = E_m$. Finally, the effective band structure can be obtained using the spectral function $A(k_i, E)$,

$$A(k_i, E) = \sum_m P_{Km}(k_i) \delta(E_m - E) \qquad (3)$$

### 5. Light Absorption

The photon energy ($\omega$) dependent absorption coefficient $\alpha(\omega)$ was calculated from real/imaginary parts of dielectric function $[\varepsilon_1(\omega)/\varepsilon_2(\omega)]$ by using VASP.

$$\alpha(\omega) = \frac{\varepsilon_2(\omega)\omega}{c \cdot \sqrt{\frac{\varepsilon_1(\omega) \pm \sqrt{\varepsilon_1^2(\omega) + \varepsilon_2^2(\omega)}}{2}}} \qquad (1)$$

The imaginary part $\varepsilon_2(\omega)$ was calculated in the random phase approximation, [92]

$$\varepsilon_2(\omega) = \frac{4\pi^2 e^2}{\Omega} \lim_{q \to 0} \frac{1}{q^2} \sum_{c,v,k} 2 w_k \delta A \qquad (2)$$

$$A = (E_{ck} - E_{vk} - \omega) \times \langle \mu_{ck+eq} | \mu_{vk} \rangle \langle \mu_{ck+e'q} | \mu_{vk} \rangle^* \qquad (3)$$

where the indices $c$ and $v$ refer to conduction and valence band states respectively, and $\mu_{ck}$ is the cell periodic part of the orbital at the k-point $\mathbf{k}$.

The real part of the dielectric tensor ($\varepsilon_1(\omega)$) is obtained by the usual Kramers-Kronig transformation:

$$\varepsilon_1(\omega) = 1 + \frac{2}{\pi} P \int_0^\infty \frac{\varepsilon_1(\omega')\omega'}{\omega'^2 - \omega^2 + i\eta} d\omega' \qquad (4)$$

Where P denotes the principle value, $\eta$ controls the complex shift.

The dense k-point meshes with grid spacing of 12x12x12 for primitive cell and 2x2x2 for the supercell (32 f.u./cell) were used for calculating ground-state band structure to guarantee that $\varepsilon_2(\omega)$ is converged. Twice the number of occupied valence bands was used for calculating empty conduction band states.

### Appendix B: Comparison of pair distribution between double-well structure and polymorphous structure

Padding improves the results at the edge of our supercell, as can be seen in Fig. 2a comparing experiment with the polymorphous calculation including padding. The agreement (black line in Fig. 2a) with experiment (red solid line) is excellent.

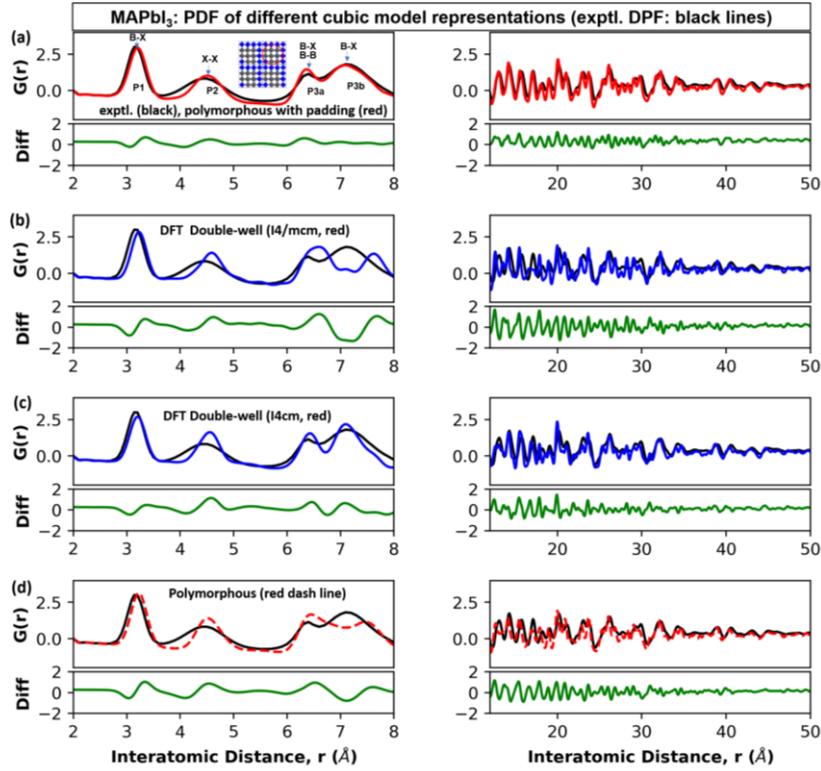

**Figure S-2.** Experimental PDF [40] (black lines) and calculated PDF from (a) polymorphous model + padding (red lines), (b,c) double-well monomorphous structures (blue lines), (d) polymorphous model without padding (red lines) of MAPbI$_3$ shown at low interatomic distances (left panels) and longer interatomic distances (right panels). All models use the same uniform $U_{ii} = 0.01$ Å$^2$ on each atom. The padded structure is shown in insert in (a). All calculations were done by using DFT-calculated periodic structures without fitted model parameters. Note that although the double well model has total energy comparable to the polymorphous 32 f.u. supercell, the PDF shows some differences, especially in peak P2 and P3. The differences between experimental and calculated PDF are depicted using green solid lines.

## Appendix C: Band gaps anomaly in monomorphous structures

### 1. Band gap order by using different exchange-correlation (XC) functionals

**Table S-3**. Band gap values ($E_g$) of orthorhombic $CsSnI_3$ and monomorphous cubic (M-cubic) $FASnI_3$ structures.

| XC functional | $E_g$ of orth-$CsSnI_3$ (4 f.u./cell) | $E_g$ of M-cubic $FASnI_3$ (1 f.u./cell) |
|---|---|---|
| PBEsol | 0.75 | 0.50 |
| PBE+ vdw | 0.83 | 0.61 |
| PBE+SOC+vdw | 0.70 | 0.39 |
| LDA | 0.75 | 0.53 |
| PBE+rVV10 | 0.74 | 0.53 |
| SCAN+rVV10 | 1.53 | 0.48 |
| HSE+SOC | 1.10 | 0.96 |
| HSE | 1.77[a] | 1.41[a] |

a: Puru Jena, J. Mater. Chem. A, 2016,4, 4728

### 2. Scheme of abnormal calculated band gaps

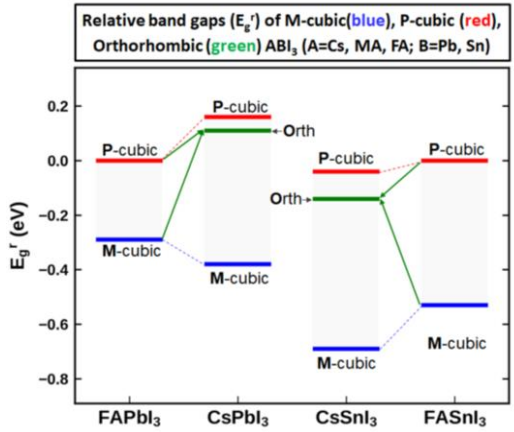

**Figure S-3.** Scheme of abnormal calculated band gap values of cubic $FASnI_3$ (Pm-3m, blue line, M-cubic) and orthorhombic $CsSnI_3$ (green line). Here we focus on the band gap difference, using $FAPbI_3$ in the polymorphous cubic (P-cubic) phase as reference. The P-cubic $FASnI_3$, $CsSnI_3$, $FAPbI_3$ and $CsPbI_3$ (red lines) are depicted in the figure, which shows the consistent gap order with the experimental observation [62].

## Appendix D: Convergence test of total energy by using cubic $CsPbI_3$

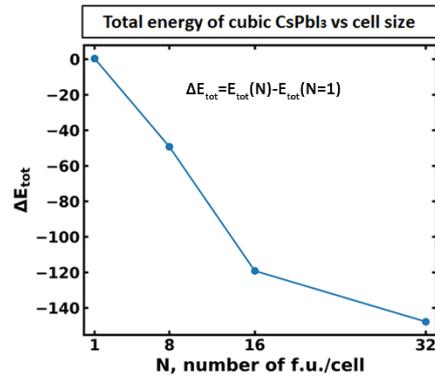

**Figure S-4.** Total energy convergence test of $CsPbI_3$ as function of cell size. This reveals that the structure with 32 f.u./cell has converged total energy.

## Appendix E: band gap renormalization

Figure S-5 shows a clear decrease in the band gap energy difference among the polymorphous and monomorphous phases when the tolerance factor increases.

## Appendix F: Distribution of octahedral deformations in halide perovskites $ABI_3$ (A=Cs, FA; B=Pb, Sn)

We calculated the distribution of octahedra deformations of monomorphous cubic phases (Figure S-6, blue solid line) with 1 f.u./cell and polymorphous cubic phases (Figure S-6, red solid line) with 32 f.u./cell for $CsPbI_3$, $CsSnI_3$, $FAPbI_3$ and $FASnI_3$. The polymorphous cubic perovskites $ABX_3$ clearly have a distribution of rotation angles (0°-15°) and B-site off-center displacement (0.0-0.75 Å). The averaged octahedra volume in polymorphous phase is also larger than that in monomorphous phase. Previous DFT calculations [40] suggested that a double-well configuration is a reasonable representation of the actual structure. The green line in Figure S-6 shows that the double-well model has a narrow distribution of motifs, thus it is a nearly monomorphous model.

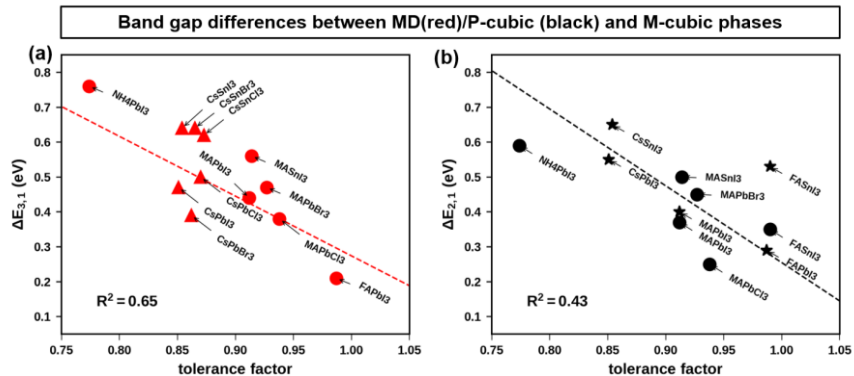

**Figure S-5.** (a) The band gap differences between MD calculations and the monomorphous cubic phases (red symbols) and (b) the difference between polymorphous and monomorphous band gaps (black symbols) as a function of tolerance factor.

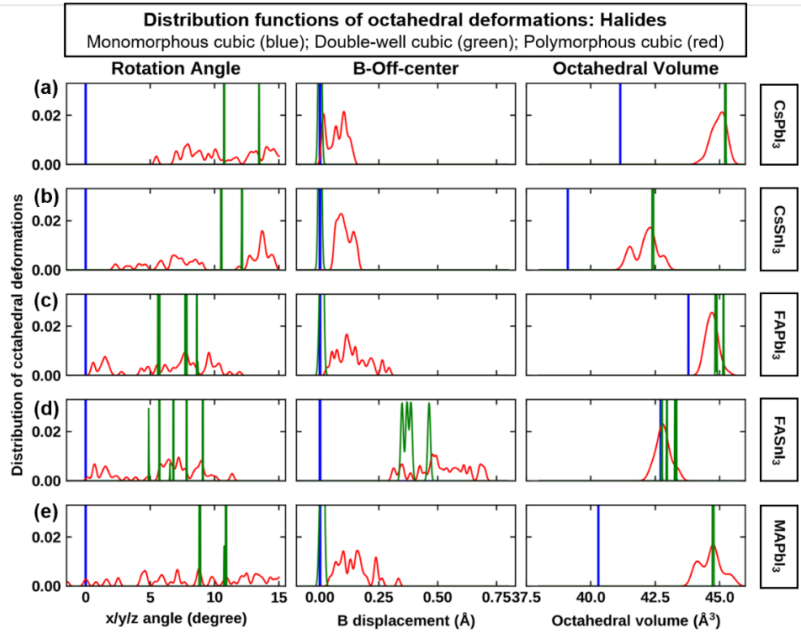

**Figure S-6.** Statistics of motifs in perovskites for (a) $CsPbI_3$, (b) $CsSnI_3$, (c) $FAPbI_3$, (d) $FASnI_3$ and (e) $MAPbI_3$. Distribution of octahedra tilting with respect to x/y/z-axis (left column), B-site off-center displacement corresponding to 6-folded X ions (middle column), and octahedral volume (right column) in polymorphous representation (The solid lines were broadened by using the Gaussian function). Blue lines are for the monomorphous cubic structure, red lines for the polymorphous cubic, and green lines for the double-well structure.

## Appendix G: Band edges compositions and shifts

### 1. Band edges orbital compositions

**Table S-7:** Orbital contribution at band edge for different $ABX_3$ perovskites in the monomorphous (M) and polymorphous (P) configurations. We show the s and p contribution to the valence band maximum (VBM) and conduction band minimum (CBM).

|   |   | VBM | | CBM | |
|---|---|---|---|---|---|
| $ABX_3$ |   | B-s | X-p | B-p | X-p |
| $CsSnI_3$ | M-cubic | 0.345 | 0.407 | 0.539 | 0.000 |
|  | P-cubic | 0.300 | 0.427 | 0.497 | 0.040 |
| $CsPbI_3$ | M-cubic | 0.245 | 0.441 | 0.564 | 0.000 |
|  | P-cubic | 0.196 | 0.462 | 0.559 | 0.035 |
| $FASnI_3$ | M-cubic | 0.309 | 0.416 | 0.523 | 0.008 |
|  | P-cubic | 0.296 | 0.424 | 0.506 | 0.026 |
| $FAPbI_3$ | M-cubic | 0.216 | 0.449 | 0.563 | 0.010 |
|  | P-cubic | 0.201 | 0.456 | 0.557 | 0.011 |

To isolate the various effects on opening the band gap, we first check the orbitals contribution at band edge for monomorphous and polymorphous structure (Table S-7). The X-p and B-s mainly contribute the VBM, and CBM mainly originates from B-p orbitals. However, there is negligible change of contribution at VBM/CBM from monomorphous to polymorphous structure. Then we create a monomorphous cell replicated periodically a few times and imposed (a) B atom displacement (0.2 Å) at constant volume, then at relaxed volume, and (b) Octahedral tilting (tilted 13 degree in plane) in fixed volume and in relaxed volume. Fig S-7 shows that both (a) B atom displacement and (b) Octahedral tilting can rise the gap in constant and relaxed volume. And B atom displacement has the larger effect.

### 2. Band edges shifts due to octahedral deformations

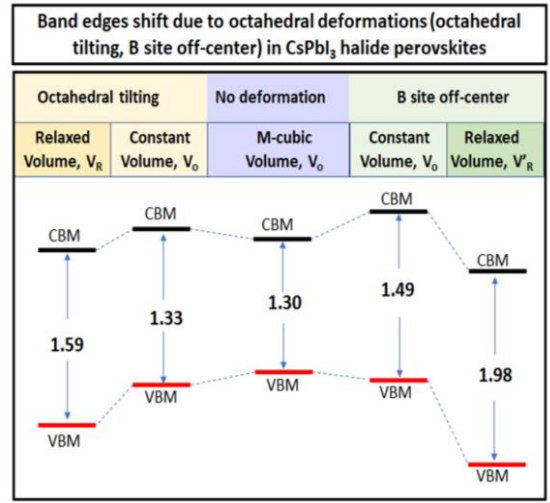

**Figure S-7.** Band edges shift (VBM, CBM) due to octahedral tilting and B-site off-center in $CsPbI_3$ halide perovskite starting from monomorphous cubic (M-cubic) $CsPbI_3$ (middle). The potential energies were aligned by using core level of I-1s orbitals in monomorphous structure.

## Appendix H: Joint density of states, transition dipole matrix and absorption

To evaluate the optical properties, we carried out the dielectric matrix calculations by using PBE functional. According to the Fermi Golden rule, the optical absorption (α) of semiconductor at photonic energy ω can be expression as follows:

$$\alpha \propto \frac{2\pi}{\hbar} \int \left| \langle v | \hat{H} | c \rangle \right|^2 \frac{1}{4\pi^3} \delta(E_c(k) - E_v(k) - \omega) d^3k$$

Where $\langle v|H|c \rangle$ is the transition matrix (i.e., oscillator strength) from states in the valence band to states in the conduction band and the integration is over the whole reciprocal space. The relative strong joint density of states (JDOS) (Fig. S-8a, red line) of polymorphous cubic FASnI$_3$ at band edge is stronger than JDOS of monomorphous cubic FASnI$_3$ (Fig. S-8a, blue line). And the comparable oscillator strength for polymorphous- (Fig. S-8b) and monomorphous- (Fig. S-8c) cubic FASnI$_3$ indicates that the polymorphous cubic FASnI$_3$ presents sharply rising absorption.

We also compare the absorption coefficient of MAPbI$_3$ derived from α=4πk/λ(Fig. 8d, black line) using experimental extinction coefficient (*k*) [93] to our calculated results (Fig. S-8d red line), using Gaussian broadening parameters (set sigma=0.3), and we find good agreement for peaks A, B, C in the range of 3.5-6.0 eV, which substantiates our model.

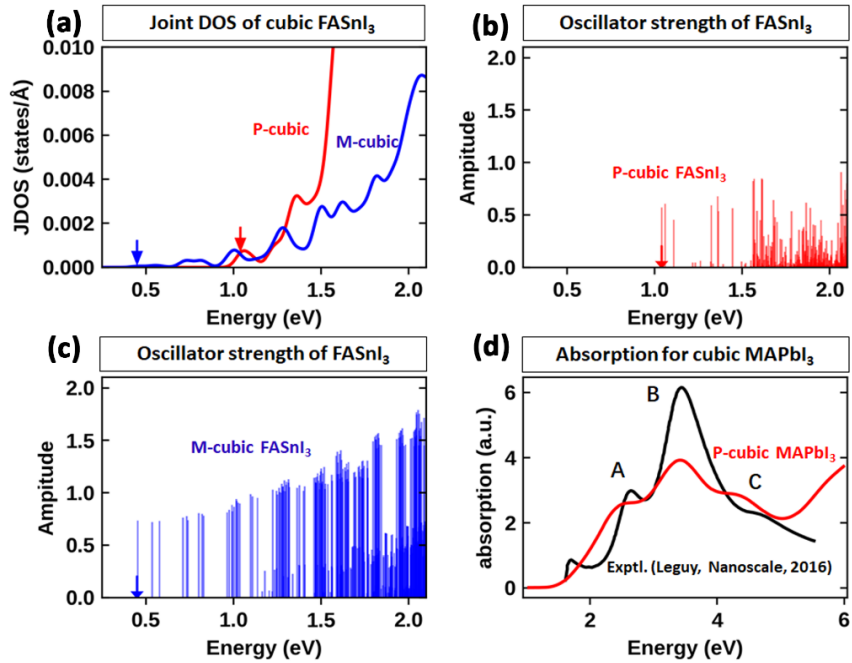

**Figure S-8:** The JDOS of the polymorphous- and monomorphous- (a) cubic FASnI$_3$. And the transition strength from valance bands to conduction bands for polymorphous- (b) and monomorphous- (c) cubic FASnI$_3$ structures. (d) Comparison of absorption between experimental absorption (solid black line) of pseudo-cubic MAPbI$_3$ [93] and calculated polymorphous cubic (solid red line, using Gaussian broadening parameters, σ=0.3) of MAPbI$_3$.